\def   \ni {\noindent}

\def   \ssk {\vskip  5truept}

\def   \bsk {\vskip 15truept}

\def   \newline {\hfil\break}

\documentstyle[epsfig]{article}
\begin{document}

\hsize 5truein
\vsize 8truein
\font\abstract=cmr8
\font\keywords=cmr8
\font\caption=cmr8
\font\references=cmr8
\font\text=cmr10
\font\affiliation=cmssi10
\font\author=cmss10
\font\mc=cmss8
\font\title=cmssbx10 scaled\magstep2
\font\alcit=cmti7 scaled\magstephalf
\font\alcin=cmr6 
\font\ita=cmti8
\font\mma=cmr8
\def\ref{\par\noindent\hangindent 15pt}
\null




\title{\ni TEV BLAZARS: STATUS OF OBSERVATIONS
}                                               

\bsk \bsk
\author{\ni  F. Krennrich$^{1}$, S.~D.~Biller$^{2}$, I.~H.~Bond$^{3}$, 
P.~J.~Boyle$^{4}$, S.~M.~Bradbury$^{3}$, A.~C.~Breslin$^{4}$,
J.~H.~Buckley$^{5}$, A.~M.~Burdett$^{3}$, J.~Bussons
Gordo$^{4}$, D.~A.~Carter-Lewis$^{1}$,
M.~Catanese$^{1}$, M.~F.~Cawley$^{6}$,
D.~J.~Fegan$^{4}$, J.~P.~Finley$^{7}$,
J.~A.~Gaidos$^{7}$, T.~Hall$^{7}$,
A.~M.~Hillas$^{3}$, R.~C.~Lamb$^{8}$, 
R.~W.~Lessard$^{7}$, C.~Masterson$^{4}$, 
J.~E.~McEnery$^{9}$, G.~Mohanty$^{1,10}$,
P.~Moriarty$^{11}$, J.~Quinn$^{12}$,
A.~J.~Rodgers$^{3}$, H.~J.~Rose$^{3}$, 
F.~W. Samuelson$^{1}$,  G.~H.~Sembroski$^{6}$,
R.~Srinivasan$^{6}$, V.~V.~Vassiliev$^{12}$,
T.~C.~Weekes$^{12}$  }                                                       
\bsk
\affiliation{ 1) Iowa State University, Physics \& Astronomy Dept.,
Ames, Ia, 50011, USA }
\affiliation{ 2) Department of Physics, Oxford University, Oxford, UK  }
\affiliation{ 3) Department of Physics, University of Leeds,
Leeds, LS2 9JT, UK}
\affiliation{ 4) Eperimental Physics Department, University College, 
Belfield, Dublin 4, Ireland  }
\affiliation{ 5)Department of Physics, Washington University, St.~Louis,
MO 63130 }
\affiliation{ 6)  Physics Department, National University of
Ireland, Maynooth, Ireland}
\affiliation{ 7) Department of Physics, Purdue University, West
Lafayette, IN 47907}
\affiliation{ 8)  Space Radiation Laboratory, California Institute of
Technology, Pasadena, CA 91125}
\affiliation{ 9) Present address: Department of Physics, University of Utah,
Salt Lake City, UT 84112}
\affiliation{ 10) Present address: LPNHE  Ecole Polytechnique, 91128 Palaiseau CEDEX, France}
\affiliation{ 11) School of Science, Galway-Mayo Institute of Technology, Galway, Ireland}

\bsk
\baselineskip = 12pt

\abstract{ABSTRACT \ni
The close relation between ground-based TeV observations
and satellite borne $\gamma$-ray measurements 
has been important for the understanding of 
blazars.  The observations which involve the
TeV component in blazar studies are reviewed.

}                                                    
\bsk
\baselineskip = 12pt

\bsk
\baselineskip = 12pt


\text{\ni 1. INTRODUCTION
\ssk
\ni 
    
Six years after the discovery of the first extragalactic $\gamma$-ray
source at TeV energies (Punch et al. 1992) the efforts to understand 
the non-thermal emission processes have stimulated the field of ground-based
$\gamma$-ray astronomy.  The experimental techniques have evolved
in a variety of aspects: starting from contemporary observations with 
several instruments covering a substantial part of the multi-wavelength
spectrum and by measuring the energy spectra and their temporal behavior.

Extragalactic TeV astronomy has been inspired by the detection of more 
than 50 AGNs between 30 MeV - 30 GeV by the EGRET detector (Thompson et al. 1995)
on-board the Compton Gamma Ray Observatory (CGRO). 
Nearly all of the EGRET AGNs are radio-loud, flat 
spectrum radio sources in which the radio emission comes
predominantly from a core region rather than from 
the outer lobes. The relatively high $\gamma$-ray luminosity of 
those sources implies that the emission is relativistically
beamed (Blandford \& Rees 1978; Blandford \& K\H onigl 1979; 
Maraschi, Ghisellini \& Celotti 1992) in the direction of the observer. 
These $\gamma$-ray AGNs with their jets aligned with the line
of sight of the observer are collectively called blazars,
but they subdivide into two classes, the flat-spectrum
radio quasars and BL Lacertae objects (review see Padovani et al. 1997).

The spectral energy distribution (SED) of the $\gamma$-ray emission
from individual blazars clearly indicates a second component 
in addition to the synchrotron spectrum at lower energies. 
The second component could be for example due to the synchrotron-self-Compton (SSC) 
mechanism as predicted by K\H onigl (1981) and Marscher \& Gear (1985).
It has also been realized, that Comptonization of external radiation
could produce $\gamma$-rays in blazars (Dermer \& Schlickeiser, \& Mastichiadis 1992;
Sikora, Begelman \& Rees 1994;  Blandford \& Levinson 1995; Ghisellini \& Madau 1996).
Also  pion photoproduction by a proton component of the jet 
(Mannheim \& Biermann 1992; Mannheim 1993)
has been suggested as source of $\gamma$-rays in blazars.

The complete coverage of the second component between MeV-multi-TeV energies
can only be achieved by using satellite-based and ground-based detectors. 
Without either one of them the spectral coverage of the second component
would be incomplete and the understanding of the emission process at 
the source significantly less constrained. 

Several ground-based $\gamma$-ray observatories (Whipple, HEGRA, CAT)
have participated in the study of blazar emission.
The Whipple observatory $\gamma$-ray telescope has started to
constrain the high energy end of the second component for 
one of the EGRET blazars: Mrk 421.
The detection of Mrk 421 (Punch et al. 1992; Petry et al. 1996)
and subsequent monitoring of this source has shown
some unexpected and extreme properties of the emission
process.

\bsk
\ni 2. VARIABILITY
\ssk
\ni 

Extreme variability on time scales ranging from  days 
(Kerrick et al. 1995) to 1/2 of one hour (Gaidos et al. 1996) 
(Figure 1) constrains the size of a presumably 
spherical emission.  

{\noindent
\parbox[t]{2.75in}{
\psfig{figure=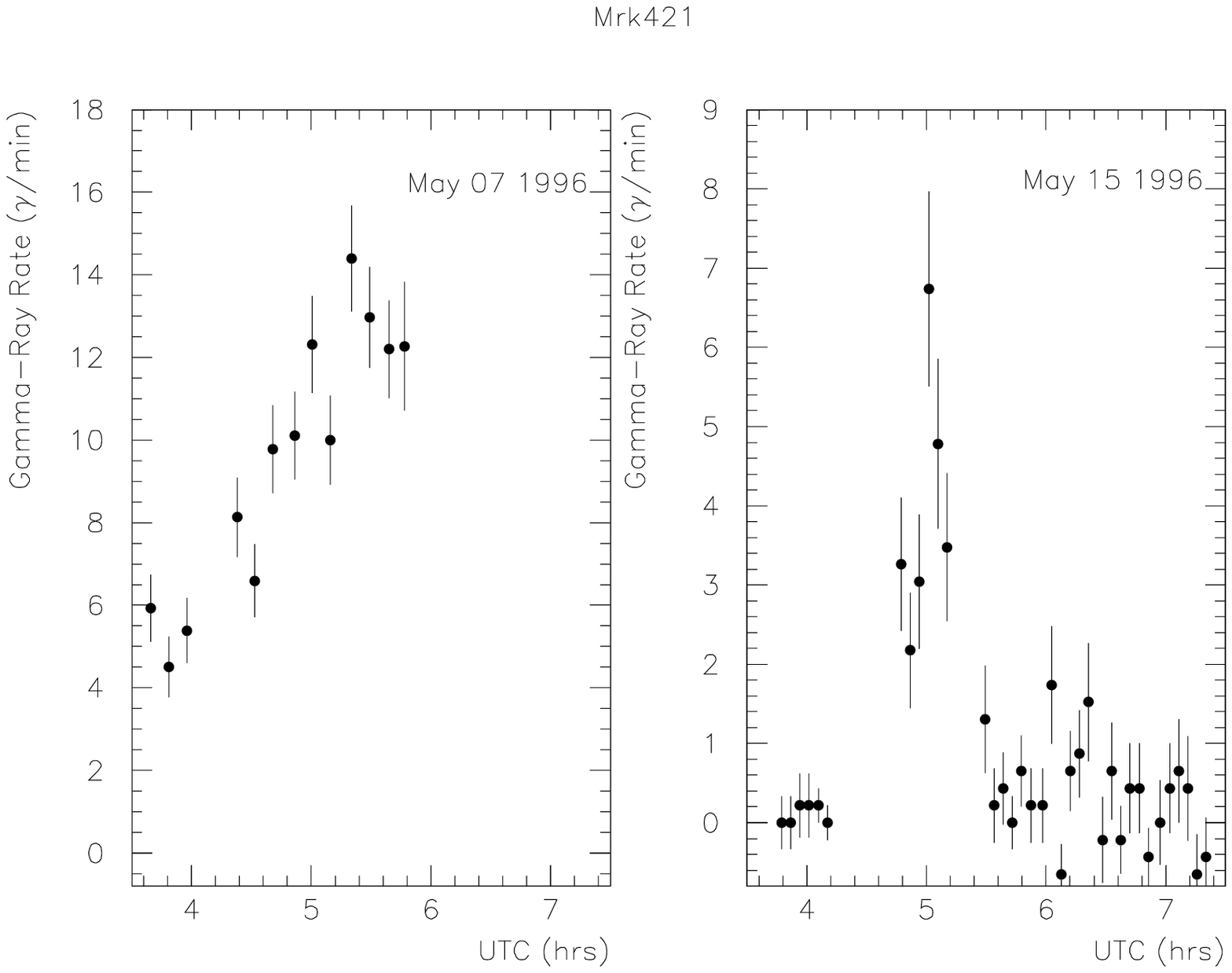,height=2.0in}
\vspace*{-0.0075in}
{\noindent {\small   \\
        \it Figure~1. Light curves of two Mrk 421 flare events of 1996 May 7 (left)
and 1996 May 15 (right). This figure has been adapted from Gaidos et al. (1996) }}}
\ \
\parbox[t]{2.15in}{
\vspace*{-1.950in}
The rapid variability implies a compact 
emission region resulting in high opacities.  To escape the 
dense emission regions the radiation has to be substantially 
Doppler boosted in order to avoid absorption through 
$\gamma$-$\gamma$ pair production. 
Assuming the TeV emission is correlated
with optical emission a Doppler factor of $\delta > 9$ has been
derived (Gaidos et al. 1996).  
For the variability of the emission from a spherical blob
relativistic causality requires that the size R of the emission
region is limited to R $\rm < c T \delta/(1+z)$.
}}

\smallskip

Gaidos et al. (1996) have shown that  $\rm R <  10^{-3}-10^{-4}  pc$.
The variability time scale places severe constraints on the geometry 
and the location of the gamma-ray emission region.  
The small emission
region might hint that TeV emission comes from close to the base of the
jet, but this cannot be firmly concluded.

\bsk
\ni 2. MULTI-WAVELENGTH OBSERVATIONS
\ssk
\ni

Studying a class of astrophysical objects with a single source provides a
limited view of the underlying physics and more sources
are needed to either derive information about them in a statistical
sense and by studying their differences.
The discovery of Mrk 501 (Quinn et al. 1996; confirmed by 
Bradbury et al. 1997) and 1ES 2344+514 (Catanese et al. 1998a) by the Whipple observatory 
$\gamma$-ray telescope (E $\rm  > 250 GeV$) has emphasized the significance of the
high energy coverage of blazars:  both, Mrk 501 and 1ES 2344+514
are not detected by EGRET. 

\bigskip

{\noindent
\parbox[t]{2.75in}{
\psfig{figure=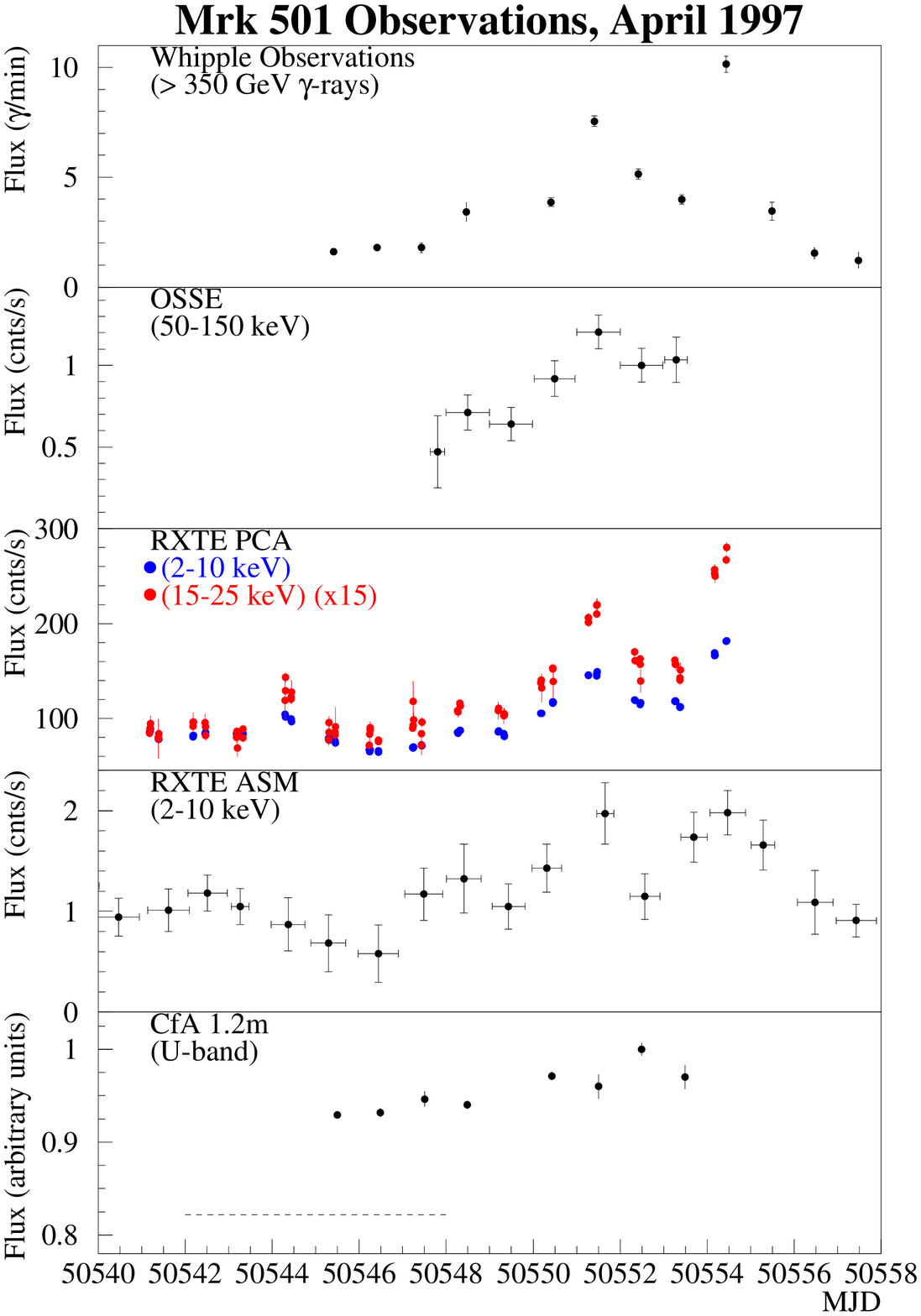,height=3.9in}
\vspace*{-0.0075in}
{\noindent {\small   \\
        \it Figure~2. Multi-wavelength observations of Mrk 501 (Catanese et al. 1998b): 
(from top to bottom): $\rm gamma$-ray, hard X-ray, soft X-ray RXTE PCA, soft X-ray RXTE ASM,
 U-band optical, taken during the period 1997 April 2-20.}}}
\ \
\parbox[t]{2.15in}{
\vspace*{-3.90in}
\\ \ \ \
Therefore, it is tempting to speculate that
some $\gamma$-ray blazars emit their maximum power at TeV energies
rather than in the  EGRET regime.
 However, simultaneous multi-wavelength
observations are necessary to study possible correlated spectral variability,
because those sources are highly variable.  Multi-wavelength campaigns
on Mrk 421 (Buckley et al. 1996) showed some evidence for correlations
between X-ray and TeV emission.
Unusual flaring activity of Mrk 501 starting in 1997 February 
(Breslin et al. 1997) lasting until 1997 August has inspired 
numerous observations including optical, X-ray, hard X-ray, MeV-GeV 
and TeV telescopes, some of them were contemporaneous and can be used
for searching for multi-wavelength correlations.
Figure 2 shows the result of one of the multi-wavelength
campaigns (Catanese et al. 1997; Pian et al. 1997) revealing possible
multi-wavelength correlations:

}}

\bigskip

Figure 2 shows the fluxes detected at different wavelengths during an
episode of flaring activity in April 1997.  The analysis of RXTE PCA
detectors (Catanese et al. 1998b) has been included in this plot.  
These new data strengthen the claim that TeV emission and X-ray
emission in the 2 keV - 25 keV band are correlated.
Several further conclusion can be derived from the multi-wavelength
campaign in April 1997:

(1) The TeV flaring activity coincides with a strong detection of
    Mrk 501 by the OSSE detector aboard the CGRO between 50 - 470 keV.
    OSSE has detected only a few blazars (McNaron-Brown et al. 1995)
    whereas Mrk 501 shows the strongest flux ever detected from a blazar
    except a high state from 3C 273 (McNaron-Brown et al. 1996).
    X-ray observations of Mrk 501 by BeppoSAX taken in April 1997 support
    that the synchrotron power in blazars can peak at hard X-ray energies
    of 100 keV (Pian et al. 1997). 

(2) The EGRET detector provides only an upper limit and this seems
    to be consistent with the interpretation that the maximum energy
    output of Mrk 501 peaks at TeV energies and not in the GeV regime,
    already suggestive from the fact that the source was
    discovered at TeV energies and not by EGRET.

(3) If the hard X-ray flux and the TeV flux are in fact correlated and this 
    were true from the optical through the $\gamma$-ray regime, one can derive 
    a lower limit on the beaming (Doppler) factor of $\rm \delta > 1.5$
    (Samuelson  et al. 1998).

Summarizing the results from the multi-wavelength campaigns it becomes evident that
there are correlations between the X-ray regime from 0.5 keV up to hard X-rays 
100 keV and the TeV emission.  
Furthermore, it shows that the synchrotron emission of Mrk 501 extends up to at least 
100 keV which is in contrast to Mrk 421 which shows a break in the spectrum at
1.6 - 2.2 keV (Takahashi et al. 1996). 
Also the data shows that the amplitude of the flux variations in the X-ray
regime are smaller than at TeV energies.

Mrk 501 has also been detected by the HEGRA experiment, the 
CAT collaboration and the telescope array (Aharonian et al. 1997; 
Djannati-Atai et al. 1997; Protheroe et al. 1997).  
Figure 3 shows the daily rates of the CAT experiment located in the
Pyrenees, the HEGRA experiment on La Palma and the Whipple observatory
in Arizona. It can be seen that the fluxes measured by the different 
telescopes are correlated.   Observations by several instruments at 
different longitudes could contribute significantly to study hourly to
daily flux variations.  
To study longterm variations over weeks and months avoiding 
gaps of several days Raubenheimer et al. (1997) have demontrated 
that atmospheric Cherenkov detectors can also be operated during 
the bright moon period. 

\begin{figure}
 \centerline{\psfig{file=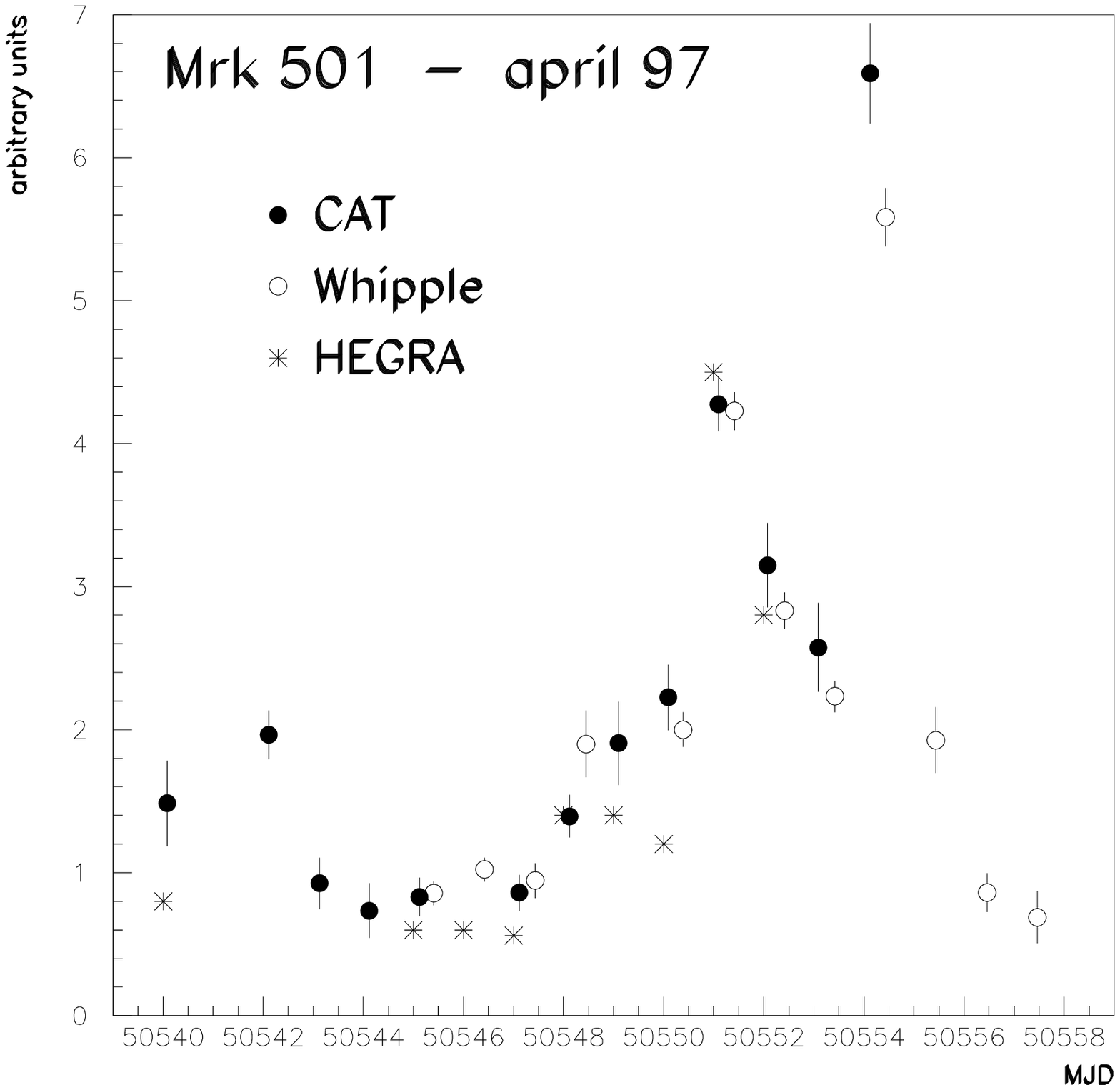, width=5.2cm}}
 \caption{FIGURE 3.  The relative fluxes of Mrk 501 in April 1997 
 measured the three different ground-based
$\gamma$-ray experiments CAT, HEGRA and Whipple (Protheroe et al. 1997).
}
\end{figure}

\bsk
\ni 2. ENERGY SPECTRA
\ssk
\ni 
The TeV energy spectra of $\gamma$-ray blazars reflect both the
physics of the $\gamma$-ray production mechanism and possibly differential
absorption effects at the source or in the intergalactic medium.
The flaring activity of Mrk 501 has provided excellent statistics to
calculate energy spectra.  Spectra have been published by several groups.
Aharonian et al. (1997) show a spectrum between 1 TeV up to 10 TeV which can 
be fit by a simple power law with a differential spectral index of 
$\rm  2.49 \pm 0.11 \pm 0.25$ (including the statistical and systematic uncertainties).

A spectrum derived over a larger energy range by Samuelson et al. (1998) (Figure~4) 
extends from 260 GeV - 12 TeV which is not well described by a simple
power-law - a simple power law fit yields $\chi^{2}$ test probability of  
 $\rm 2.8 \times 10^{-7}$.  
Instead the spectrum exhibits significant curvature and can be well fit
by a parabolic spectrum:

$ \rm J(E) = (8.6 \pm 0.3 \pm 0.7) \: \times \: 10^{-7} \: E^{-2.22
\: \pm \: 0.04 \pm \: 0.05 \: -(0.47 \pm 0.07)log_{10}(E) } 
\: m^{-2}  s^{-1}  TeV^{-1} $.

Note, that there is not necessarily a contradiction between the two different
spectra derived since Aharonian et al. (1997) cover a smaller energy range 
and give larger uncertainties in the spectral index.

\begin{figure}
 \centerline{\psfig{file=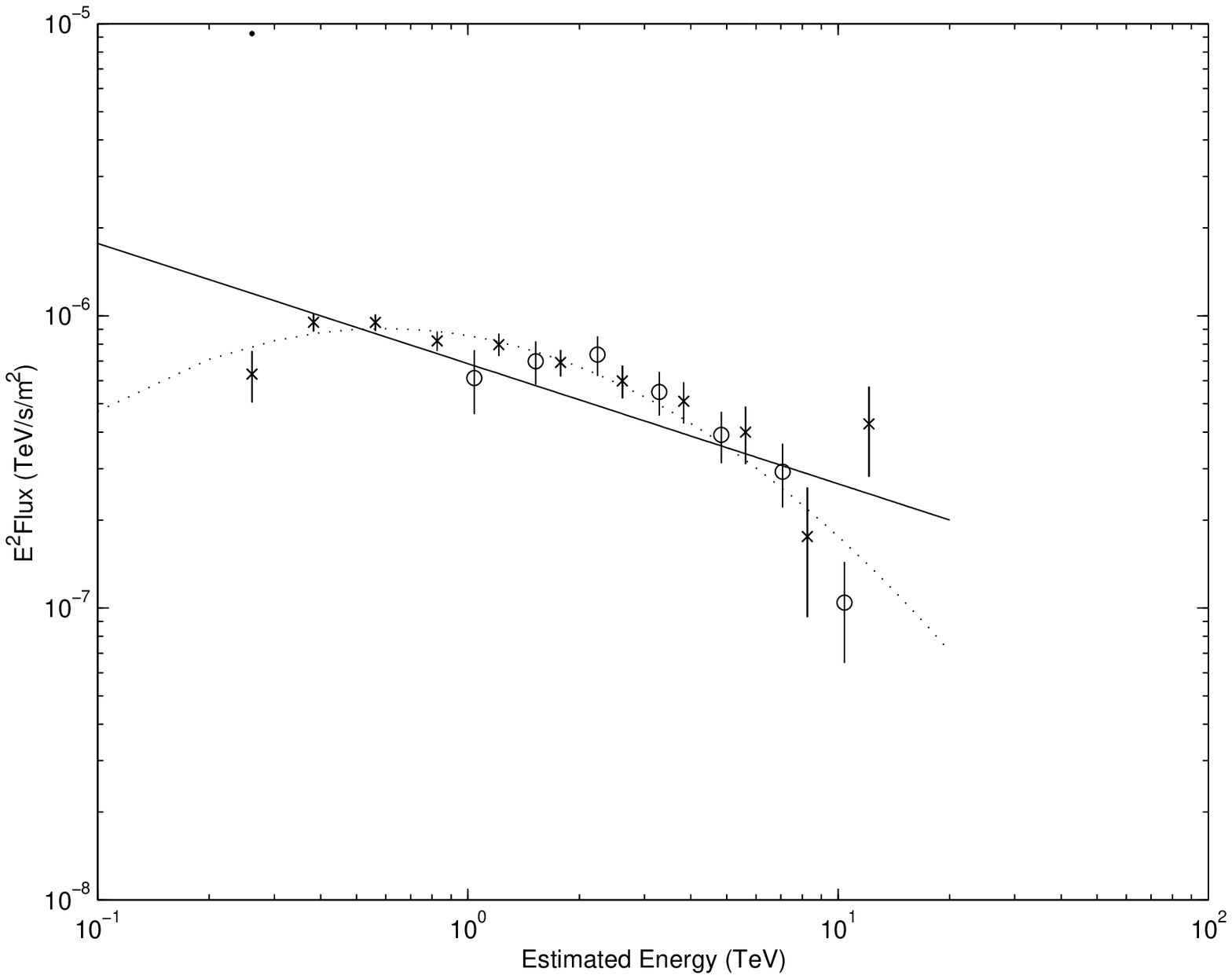, width=7.2cm}}
 \caption{FIGURE 4.  The energy spectral distribution of Mrk 501
between 260 GeV - 12 TeV, measured with the Whipple 10m gamma-ray telescope
(Samuelson et al. 1998). The stars show the differential fluxes 
derived from 15 hours of observations at small (standard) zenith angles,
the circles show the fluxes derived from large zenith angle observations
(details given in Krennrich et al. 1999).
}
\end{figure}

Possible causes of curvature at the lower energies could be related to the
 intrinsic spectrum of the source requiring that the spectrum has to
be consistent with the EGRET upper limits. 
A small subset of (Samuelson et al. 1998) nearly contemporaneous TeV/GeV
observations show that a curved spectrum is consistent with the EGRET
upper limits whereas a simple power law is not.
Causes of curvature at higher energies include partial absorption
of $\gamma$ rays in the intergalactic medium, absorption in the jet by pair
production on low energy photons, absorption by photons near the source, 
or an end in the primary particle beam energy. 
Intergalactic absorption by infrared photons has been advocated by a number of
authors (Gould, R. J. \& Schr\` eder G. 1967; Stecker \& De Jager 1992; 
Dwek \& Slavin 1994; Biller et al. 1995) as a potential cause of such absorption.
Therefore, the oberved spectrum might be a convolution of the blazar spectrum 
with an external absorption term. By considering the extent of this potential
modification of the blazar spectrum, upper limits to the density of the
intergalactic IR field may be determined.  Such limits have been recently derived by
Biller et al. (1998) in a robust manner using data from Mrk 421 and Mrk 501. 
The curved spectrum of Mrk 501 (Samuelson et al. 1998) can still accomodate the 
limits from Biller et al. (1998).

The energy spectrum of Mrk 421 (Krennrich et al. 1999) based on short flares 
has been derived with a similar statistical precision 
as the spectrum of Mrk 501 in Samuelson et al. (1998). The resulting spectrum 
of Mrk 421 is fit reasonably well by a simple power law; 
 $ \rm J(E) = \: \:
E^{-2.54 \: \pm \: 0.03 \: \pm \: 0.10} \: photons \: m^{-2} \: s^{-1}
\: TeV^{-1} $

This is in contrast to the spectrum of Mrk 501. In figure 5 the spectra of
Mrk 501 and Mrk 421 are shown in comparison. The spectra are different 
in shape and since Mrk 421 and Mrk 501 have almost the same redshift
(0.031 and 0.033 respectively), the difference in their spectra must be
intrinsic to the sources and not due to intergalactic absorption,
assuming the intergalactic infrared background is uniform.

At GeV energies, Mrk~421 is seen by EGRET (Lin et al. 1992),
albeit weakly, whereas Mrk~501 is not (Catanese et al. 1997). For the
latter it could be argued that in
a synchrotron-inverse Compton picture it would appear that
both the synchrotron and the inverse-Compton peak are shifted to
higher energies leaving the EGRET GeV energy sensitivity range in the
gap between the peaks.  As shown in Fig.~5, in the energy
range 260 GeV~-~10~TeV, the spectrum of Mrk~501 is harder at lower
energies and shows more curvature than Mrk~421. In fact the latter can 
be fit by a straight line (i.e., pure power-law).  This is
also consistent with the peak inverse-Compton power occuring at higher
energies for Mrk~501. 
Therefore, also the detailed energy spectra of Mrk 421 and Mrk 501
concur well with the  conclusion drawn from the multi-wavelength
spectrum: in case of Mrk 501 the TeV emission is closer to the 
maximum power peak of the second component than in case of Mrk 421.
Blazar emission can also reveal important information about the 
underlying source mechanism through spectral variability.
Short term flare spectra of Mrk 421 have been derived in 
Krennrich et al. (1999).  
Figure 6 shows the spectra of
Mrk 421 during several flares: a big flare (2 hours: 7.4 Crab) on May 7 1996 (flare I), 
a flare (30 minutes: 2.8 Crab) on May 15 1996 (flare II) and flares from 3 different
nights (1.5 hours total: 3.3 Crab) in June 1995 (flare III). 

{\noindent
\parbox[t]{2.450in}{
\psfig{figure=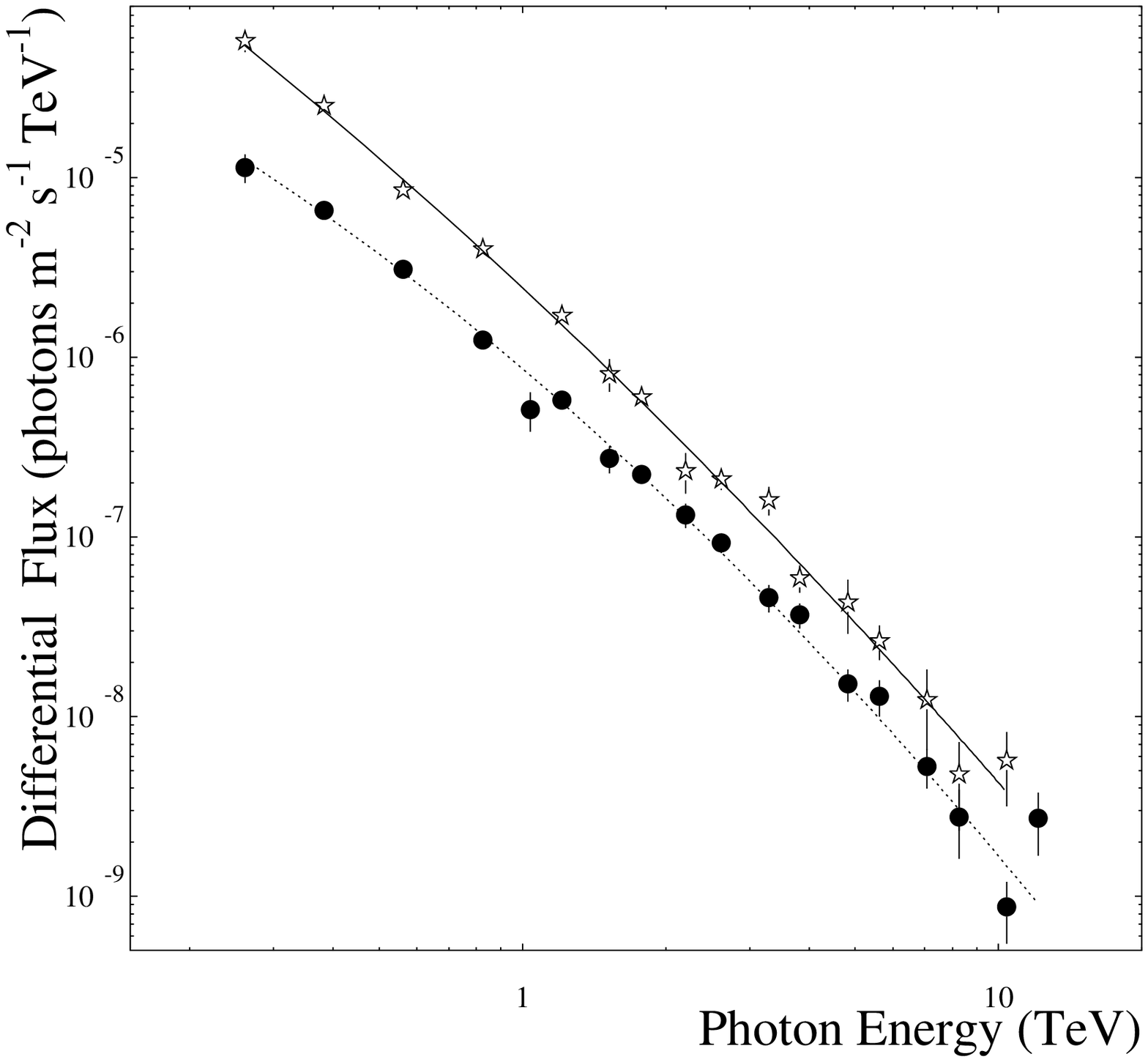,height=2.75in}
\vspace*{-0.000075in}
{\noindent {\small   \\
\ \
        \it FIGURE 5. The energy spectral distribution of Mrk 501
between 260 GeV-12 TeV (filled circles) compared to the spectrum of Mrk 421 (open stars)
(Krennrich et al. 1999) }}}
\ \
\parbox[t]{2.450in}{
\psfig{figure=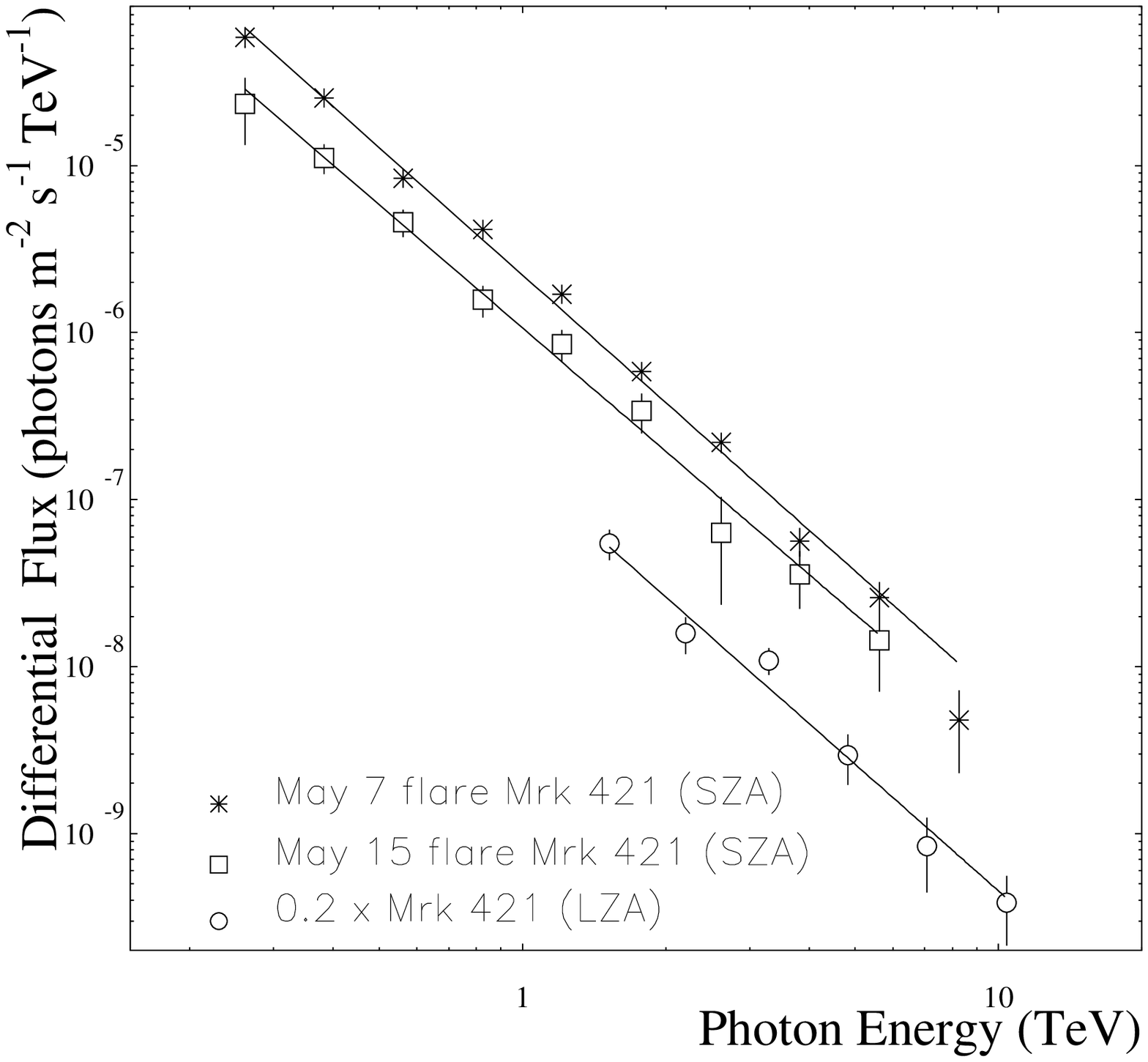,height=2.75in}
\vspace*{-0.000075in}
{\noindent {\small  \\
          \it FIGURE 6.  The energy spectra of Mrk 421 during different flaring states (Krennrich et al. 1999): 
``Big Flare'' (7.4 Crab) on May 7 1996, flare (3.3 Crab) from May 15 1996 and a flare (2.8 Crab) 
at large zenith angles from June 20, 29 and 30 1995 }
}}

\smallskip
\smallskip
\smallskip

  The average fluxes (Crab units) differ by a factor of 2.6 
and the signature of the flares are different: flare I shows a clear rise,
whereas flare II includes the rise and the fall and flare III show a
constant emission.  
The energy spectra show the same shape and spectral index for all three
flares (Krennrich et al. 1999).  There is no significant change. 
Changes in the spectral index with the flux have also been studied in
great detail by (Aharonian et al. 1998) using Mrk 501 observations from the HEGRA 
array of telescopes.
Also, no significant variations of the spectral index are apparent in
the data.

\bsk
\baselineskip = 12pt
{\abstract \ni ACKNOWLEDGMENTS  The invitation to the 3rd Integral workshop
is greatly appreciated. 

}

\bsk
\baselineskip = 12pt


{\references \ni REFERENCES
\ssk

\ref Aharonian, F. A., et al.
1997, A \& A, {\bf 327}, L5

\ref Aharonian, F. A., et al.
1998, astro-ph/9808296

\ref Biller, S. D., et al. 1998, Phys. Rev. Lett., {\bf 80}, 2992

\ref  Blandford, R. D. \& K\H onigl, A. 1979, ApJ,  {\bf 232}, 34

\ref Blandford, R. D. \& Rees, M. J. 1978, in Pittsburg Conf. on BL Lac Objects,
                          ed. A. N. Wolfe (Pittsburg University Press), 328

\ref Blandford, R. D. \& Levinson, A.,  1995, ApJ,{\bf 441}, 79

\ref Bradbury, S. M., et al. 1997, A \& A,{\bf 320}, L5

\ref Breslin. A. C. et al. 1997, IAU Circ. 6596

\ref Buckley, J. H., et al. 1996, ApJ, {\bf 472}, L9

\ref Catanese, M., et al. 1997, ApJ,{\bf 487}, L143

\ref Catanese, M., et al. 1998, ApJ, {\bf 501}, 616

\ref Catanese, M., et al. 1998, Conf. on BL Lac Phenomenon, Turku (Finland), in press

\ref Dwek, E., \& Slavin, J. 1994, ApJ., {\bf 436}, 696

\ref Djannati-Atai, A. et al. 1997,   Proc. of Towards 
                      a Major Atmospheric Cherenkov Detector-V,
                      ed. by O. C. De Jager, 21

\ref Dermer, C. D., Schlickeiser, R., \& Mastichiadis 1992, A., A \& A, {\bf 256}, L27

\ref Gaidos, J. A., et al. 1996,
 Nature,  {\bf 383}, 319

\ref Ghisellini, G. \& Madau, P. 1996, MNRAS, {\bf 280}, 67

\ref Gould, R. J. \& Schr\` eder G. 1967,
Phys. Rev.,  {\bf 155}, 1408

\ref Kerrick, A. D., et al. 1995, ApJ, {\bf 438}, L59

\ref  K\H onigl, A., 1981, ApJ, {\bf 243}, 700

\ref Krennrich, F., et al. 1997, ApJ, {\bf 482}, 758

\ref Krennrich, F., et al. 1999, ApJ, {\bf 511}, in press

\ref Lin, Y. C., et al. 1992, ApJ, {\bf 390}, L49

\ref Mannheim, K. \& Biermann, P. L., A \& A, {\bf 253}, L21

\ref Mannheim, K. 1993,  A \& A, {\bf 269}, 67

\ref Maraschi, L., Ghisellini, G., \& Celotti, A. 1992, ApJ,  {\bf 397}, L5

\ref Marscher, A.,  \& Gear, W. 1985, ApJ, {\bf 198}, 114

\ref McNaron-Brown, K. et al. 1995, ApJ, {\bf 451}, 575 

\ref McNaron-Brown, K. et al. 1996, ApJ, {\bf 474}, L85 

\ref Mattox, J. R. et al. 1993, {\bf 410}, 609

\ref Padovani, P., 1997, Proc. of XXXIInd Rencontres de Moriond, Les Arcs, Savoie, 
                         ed. by Giraud-Heraud \& Tran Thanh Van, 7

\ref Petry, D. et al. 1996, A \& A,{\bf 311}, L13  

\ref Pian, E. et al. 1997, ApJ, {\bf  492}, L17

\ref Protheroe, R. J. et al. 1997, Proc. of the 25th ICRC (Durban, South Africa), 8, 317

\ref Punch. M., et al. 1992, Nature, {\bf 358}, 477

\ref Quinn, J., et al. 1996, ApJ, {\bf 456}, L63

\ref Samuelson, F. W. 1998, ApJ, {\bf 501}, L17

\ref Sikora, M., Begelman, M. C., \& Rees, M. J. 1994, ApJ, {\bf 421}, 153

\ref Stanev, T.  \& Franceshini, A. 1998, ApJ, {\bf 494}, L159

\ref Stecker, F. W., De Jager, O. C. \&  Salamon, M. H.  1992, ApJ, {\bf 390}, L49 

\ref Takahashi, T. et al. 1996, ApJ, {\bf 470} , L89

\ref Thompson, D. J. et al. 1995, ApJ Suppl., {\bf 101}, 259

}                      

\end{document}